\documentclass[twocolumn,prl,showpacs,amsfonts,superscriptaddress,floatfix,nofootinbib]{revtex4-1}
\pdfoutput=1
\usepackage{amsmath}
\usepackage{graphicx}
\usepackage{footmisc}
\usepackage{color}

\graphicspath{ {fig2/} }

\begin{document}
\title {Spontaneous breaking of time-reversal symmetry in topological insulators}
\author{Igor N.Karnaukhov}
\affiliation{G.V. Kurdyumov Institute for Metal Physics, 36 Vernadsky Boulevard, 03142 Kiev, Ukraine}

\begin{abstract}
{
The system of spinless fermions on a hexagonal lattice is studied . We have considered tight-binding model with the hopping integrals  between the nearest-neighbor and next-nearest-neighbor lattice sites, that depend on the direction of the link. The links are divided on three types depending on the direction, the hopping integrals are defined by different phases along the links.
The energy of the system depends on the phase differences, the solutions for the phases, that correspond to the minimums of the energy, lead to a topological insulator state with the nontrivial Chern numbers.
We have analyzed distinct topological states and phase transitions, the behavior of the chiral gapless edge modes, have defined the Chern numbers. The band structure of topological insulator (TI) is calculated, the ground-state phase diagram in the parameter space is obtained. We propose a novel mechanism of realization of TI, when the TI state is result of spontaneous breaking of time-reversal symmetry due to nontrivial stable solutions for the phases that determine the hopping integrals along the links and show that the
Haldane model \cite{Hal} can be implemented in real compounds of the condensed matter physics.
}
\end{abstract}

\maketitle
\section{Introduction}
Topological phases of matter such as  TIs, topological supercunductors, fractional topological insulators have gained considerable attention \cite{Hal,10,9,KM}.
TIs are characterized by the topological invariants and a Kramers pair of gapless helical edge states.
An artificial magnetic field in the Haldane model \cite{Hal} or an external magnetic field in the Hofstadter model \cite{Hof} break time-reversal symmetry and lead to nontrivial topological state of the system. The breaking symmetry is a result of the inclusion of the spin-orbit interaction \cite{KM}, vector potential, that defines the Peierls phase \cite{Hal, Hof}, $(d+id)$ (or $(p+ip)$) superconducting order parameter \cite{TSC5,TSC5a,TSC6} in the model Hamiltonians.
Ultra-cold atoms trapped in optical lattices represent the platform for the potential realization TIs \cite{0},\cite{ref}, the
Haldane  model \cite{Hal}. The Hofstadter model has been realized experimentally in ultracold atoms using laser assisted tunnelling in tilted optical lattices \cite{Hof1}. The Haldane TI can not be  realized in real compounds of the condensed matter physics.

The standard method of realization of topological states is reduced to the consideration of symmetry breaking of the system due to the presence of additional interaction. The topological state are described in the framework of topological band theories \cite{10,9, Hal, KM} (as a rule without interaction), which characterize a class of TI by a topological invariant. The simplest models of the quantum spin Hall effect \cite{a1,a2,a3} can be
understood as direct products of time-reversed copies of a Chern insulator, this picture was generalized to the ${\mathrm{Z}}_2$ topological band insulators \cite{b1,b2,b3,b4,b5}. The presence of time reversal and the broken spin rotational symmetry has been discussed for three dimensional class DIII topological superconductors in \cite{R}.
We consider another nature of the realization of TI, when symmetry is spontaneously broken by the stable thermodynamic state, the state of the system selects a global difference phases of the hopping integrals of fermions.

The hopping integrals $t_{ij}$ are defined by the overlap of the wave functions of the fermions localized at the lattice sites $i,j$ $t_{ij}\sim \int d\textbf{r} \psi^\ast (r-r_i)\psi(r-r_j)$. The wave function  $\psi(r)$ is the solution of the Schr\"{o}dinger equation, at the same time $\psi(r-r_j)$ is defined with accuracy of the phase $\chi_j$. There is no absolute value for the phase of the wave function: any phase can be chosen between 0 and $2\pi$. Redefine the hopping integral explicitly highlighting phase  $t_{ij} = t_{|i-j|}\exp(i (\chi_j -\chi_i))$, where $t_{|i-j|}\sim \int d\textbf{r} |\psi (r-r_i)| |\psi(r-r_j)|$. The phase difference may determine the physical properties of the system as in the Josephson effect, where a dissipationless current across a junction between two superconductors defines by the phase difference of the superconducting order parameters. We consider another possibility of realization of the physical properties of the system due to the specified phases of the wave function.

The absence of any required interaction or external field is of course a experimental simplification in the investigation of topological states. In the framework of the proposed approach we show that the Haldane TI can be realized in 2D materials. Using a simple tight-binding model of spinless fermions we show that there is family of models that describe a stable TI state due to spontaneous breaking of time-reversal symmetry. In other words the continuous gauge symmetry is spontaneously broken by the stable thermodynamic state, the state of the system selects global difference phases that define topological order parameter. In this paper we address the behavior of a tight-binding model defined on the hexagonal lattice with emphasis on topological phase states that are specific to considered lattice.
\section{The model}

We describe a very intuitive scheme of construction of the new family of models with nontrivial topological
properties using the tight-binding model of spinless fermions. We will analyze the behavior of spinless fermions on a hexagonal (honeycomb) lattice in the framework of the  single-particle Hamiltonian describing the hoppings of the spinless fermions between nearest- and next-nearest neighbor lattices sites
\begin{multline}
    {\cal H}= - \sum_{\alpha=x,y,z}\sum_{\alpha-links}t_{ij}^\alpha a^\dagger_{i}a_j     -\sum_{\beta=X,Y,Z} \sum_{\beta-links}\tau_{ij}^\beta a^\dagger_{i}a_j \\
    -h\sum_{j=A} a^\dagger_{j}a_j + h\sum_{j=B} a^\dagger_{j}a_j ,
        \label{eq-H}
\end{multline}
where $a^\dagger_{j} $ and $a_{j}$ are the spinless fermion operators on a site \emph{j} with the usual anticommutation relations. The Hamiltonian describes the hoppings of fermions between the nearest-neighbor lattice sites with the magnitudes $t_{ij}^{\alpha}=t\exp(i \phi_{ij}^\alpha)$ $\alpha =x,y,z$ and the next-nearest-neighbor lattice sites with the magnitudes  $\tau_{ij}^{\beta}= \tau \exp (i \varphi_{ij}^\beta)$ $\beta =X,Y,Z$, with $\varphi_{ij}^\beta = \varphi_{A}^\beta(ij)$ for hoppings of the fermions located at the A-atoms and $\varphi_{ij}^\beta = \varphi_{B}^\beta (ij)$ at the B-atoms, $t$ is equal to unit, $h$ is a staggered potential. The hopping integrals are differed the phases along the links $x,y,z$  between the nearest-neighbor and along the links $X,Y,Z$  between the next-nearest-neighbor lattice sites.

\begin{figure}[tbp]
    \centering{\leavevmode}
    \centering{\leavevmode}
\begin{minipage}[h]{.4\linewidth}
\center{
            \includegraphics[width=\linewidth]{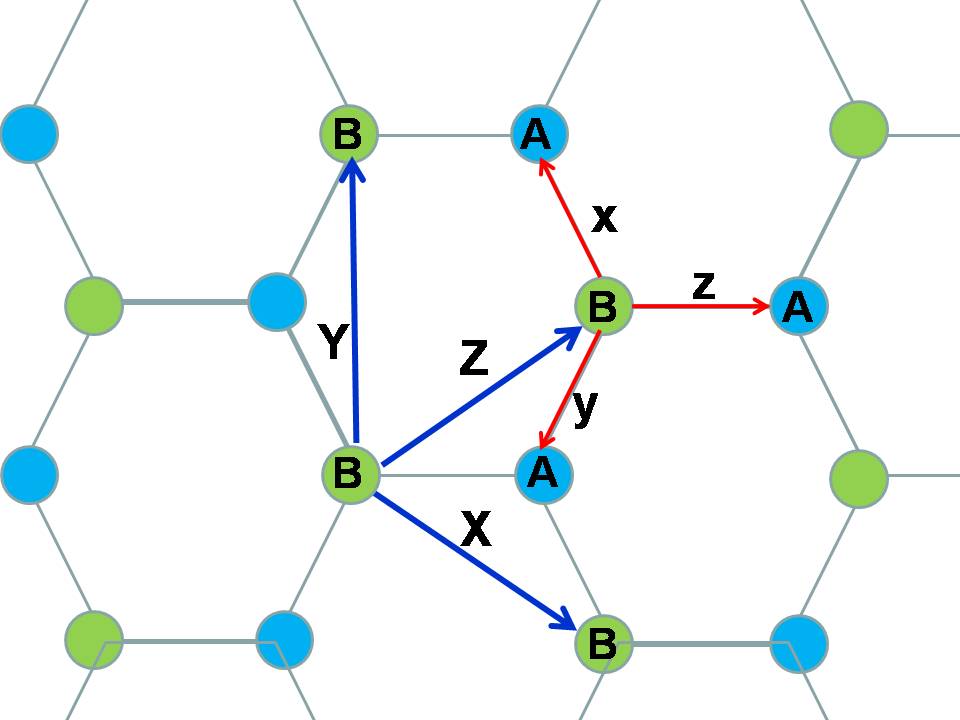}
                  }
    \end{minipage}
    \caption{(Color online)
       Hexagonal (honeycomb) lattice, tree type of links x-,y-,z- with hopping integrals  $t_{\alpha}=t\exp(i \phi^\alpha)$ $\alpha =x,y,z$ for the fermions located at the A- and B-atoms (nearest-neighbors) and the links X-,Y-,Z- with hopping integrals $\tau_{A,B}^{\beta}= \tau \exp (i \varphi_{A,B}^\beta)$ $\beta =X,Y,Z$ for fermions located at the atoms of the same type (next-nearest-neighbors), the unit cell contains two atoms A and B.
 } \label{fig:0}
\end{figure}

We will define the phases of the hopping  integrals in unit cell (see in Fig.\ref{fig:0})
$\phi_{ij}^\alpha =\phi^\alpha$ and $\varphi_{A,B}^\beta(ij)=\varphi_{A,B}^\beta$
and solve the model analytically for the uniform configuration, due to the translational invariance of the lattice. As a rule, the ground-state energy of different systems is minimized by the same uniform configurations of the gauge fields \cite{Kitaev, Lieb}, in our case with considered static configurations.

We can continuously deform the hexagonal lattice changing the angle between the translation vectors to $\frac{\pi}{2}$, using these perpendicular vectors as a basis of the translation group. The behavior of the system is defined by stable solutions for the phase differences, we can reduce the number of phases with the help of local canonical transformationû for fermions operators. Shifting the $k_x,k_y$-components of the wave vector and considering the phase differences we can decrease the number of unknown phases to six $\varphi_{A,B}^{x,y,z}$. The model (\ref{eq-H}) is defined by two parameters, $\tau$ and $h$ are free parameters in the model Hamiltonian, they are defined in the units of the hopping integral $t$.  We ignore spin of the particles and focus on the relatively simpler case of TI with spontaneously broken time-reversal symmetry.

\section{Topological insulator state}

\begin{figure}[tp]
    \centering{\leavevmode}
\begin{minipage}[h]{.4\linewidth}
\center{
           \includegraphics[width=\linewidth]{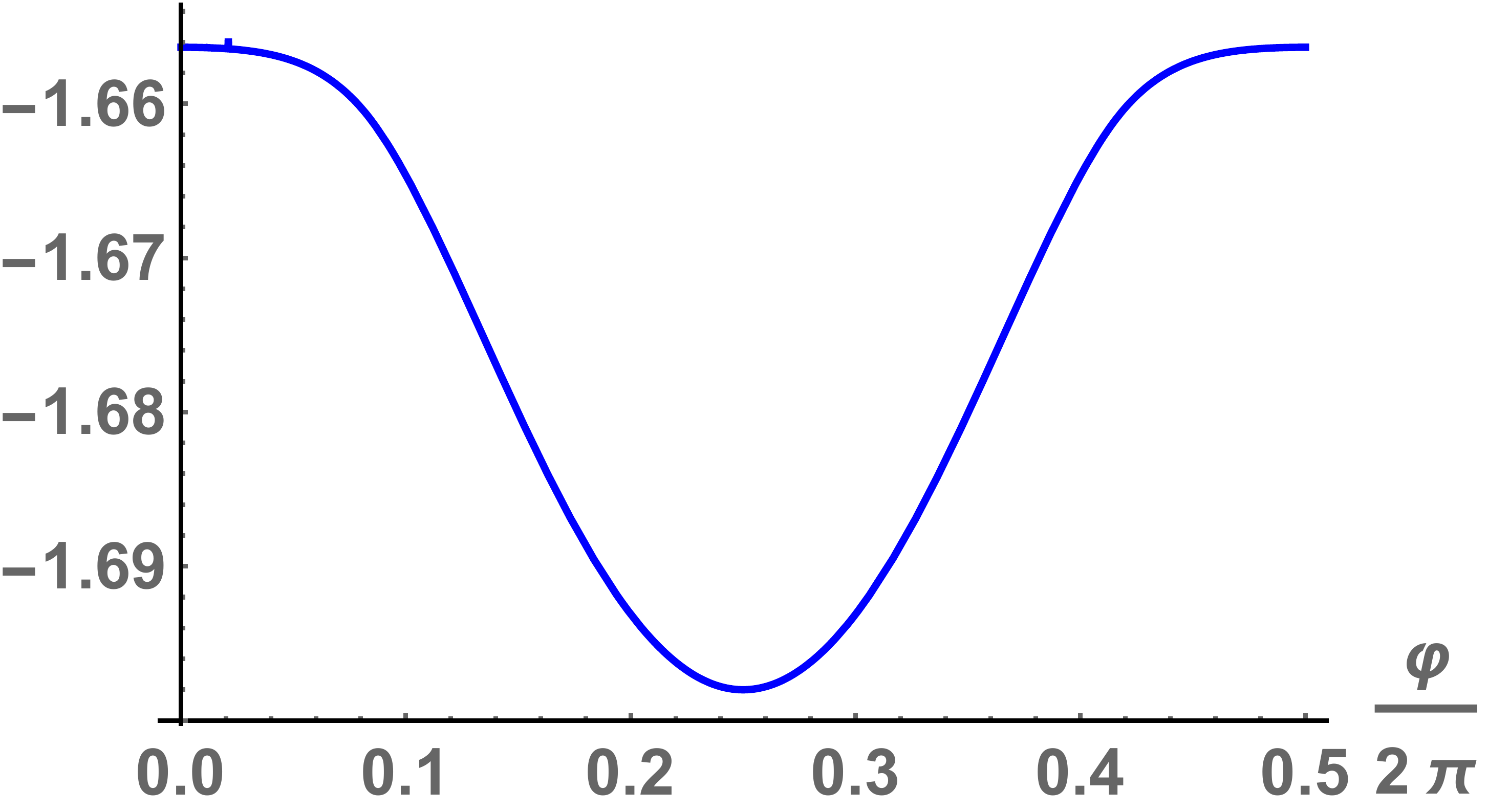} \\ a)
                  }
    \end{minipage}
       \centering{\leavevmode}
\begin{minipage}[h]{.5\linewidth}
\center{
           \includegraphics[width=\linewidth]{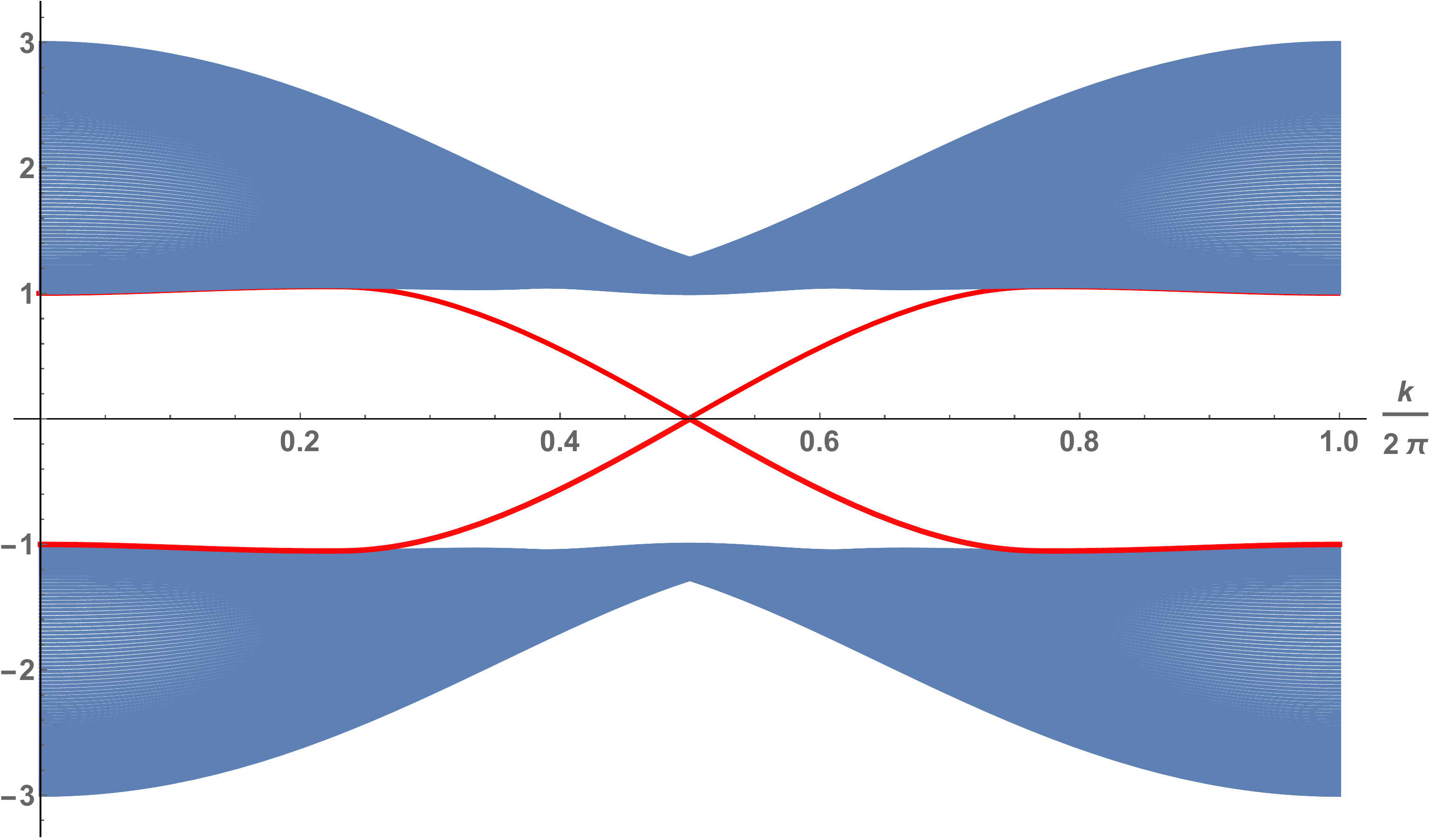}\\ b)
                  }
    \end{minipage}
    \begin{minipage}[h]{.4\linewidth}
\center{
           \includegraphics[width=\linewidth]{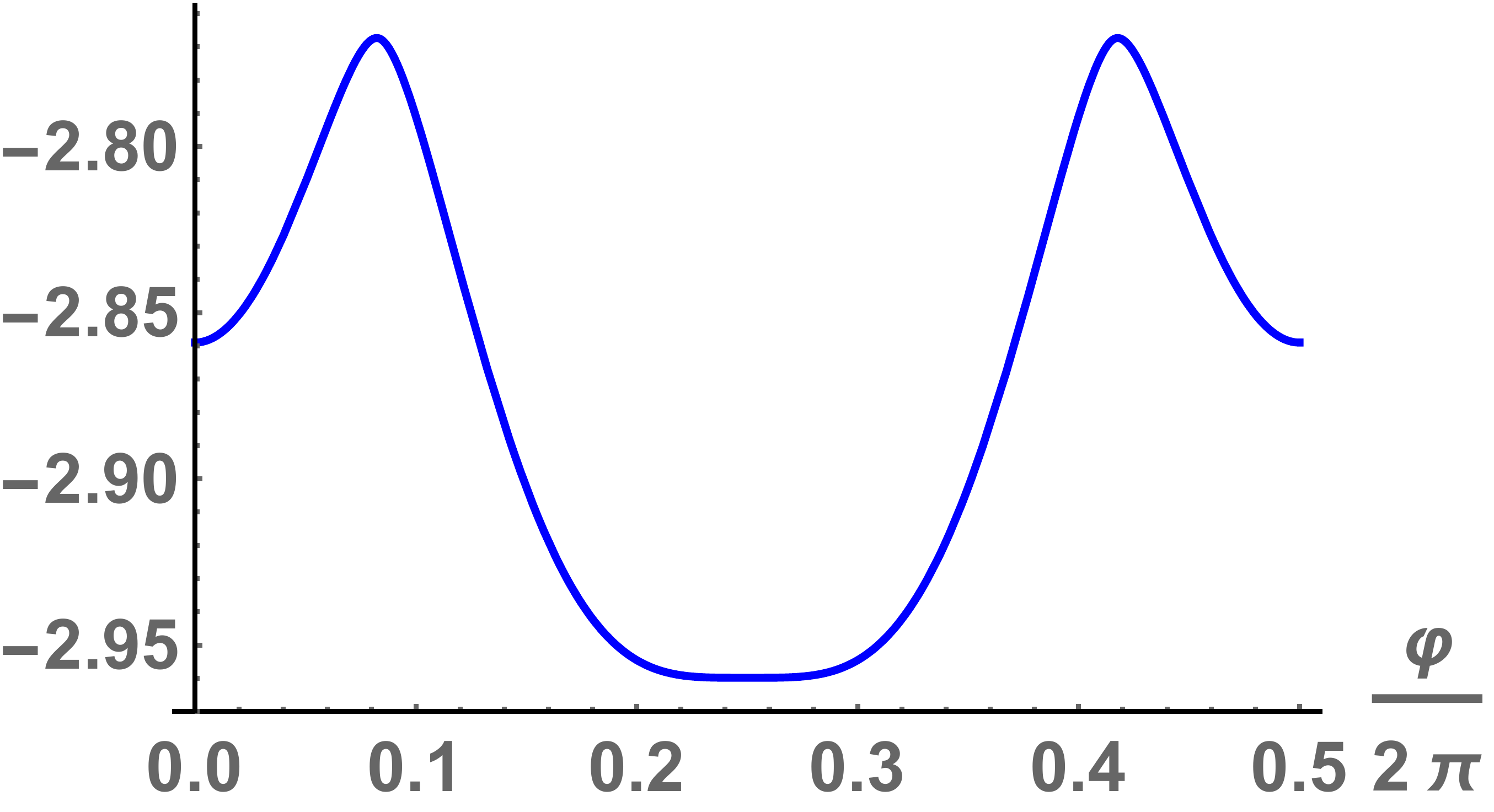}\\ c)
                  }
    \end{minipage}
    \begin{minipage}[h]{.5\linewidth}
\center{
   \includegraphics[width=\linewidth]{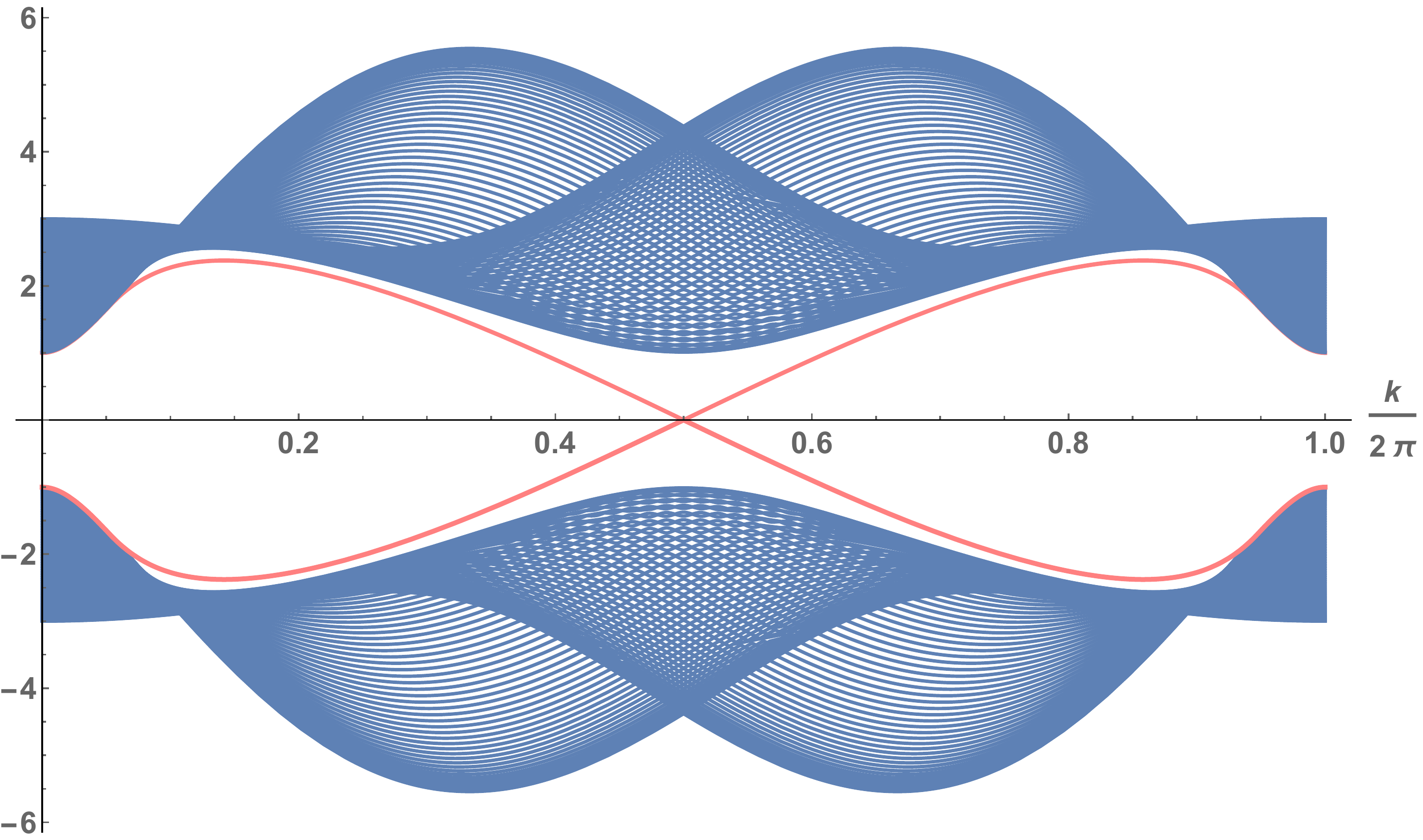}\\ d)
                  }
    \end{minipage}
      \begin{minipage}[h]{.4\linewidth}
\center{
          \includegraphics[width=\linewidth]{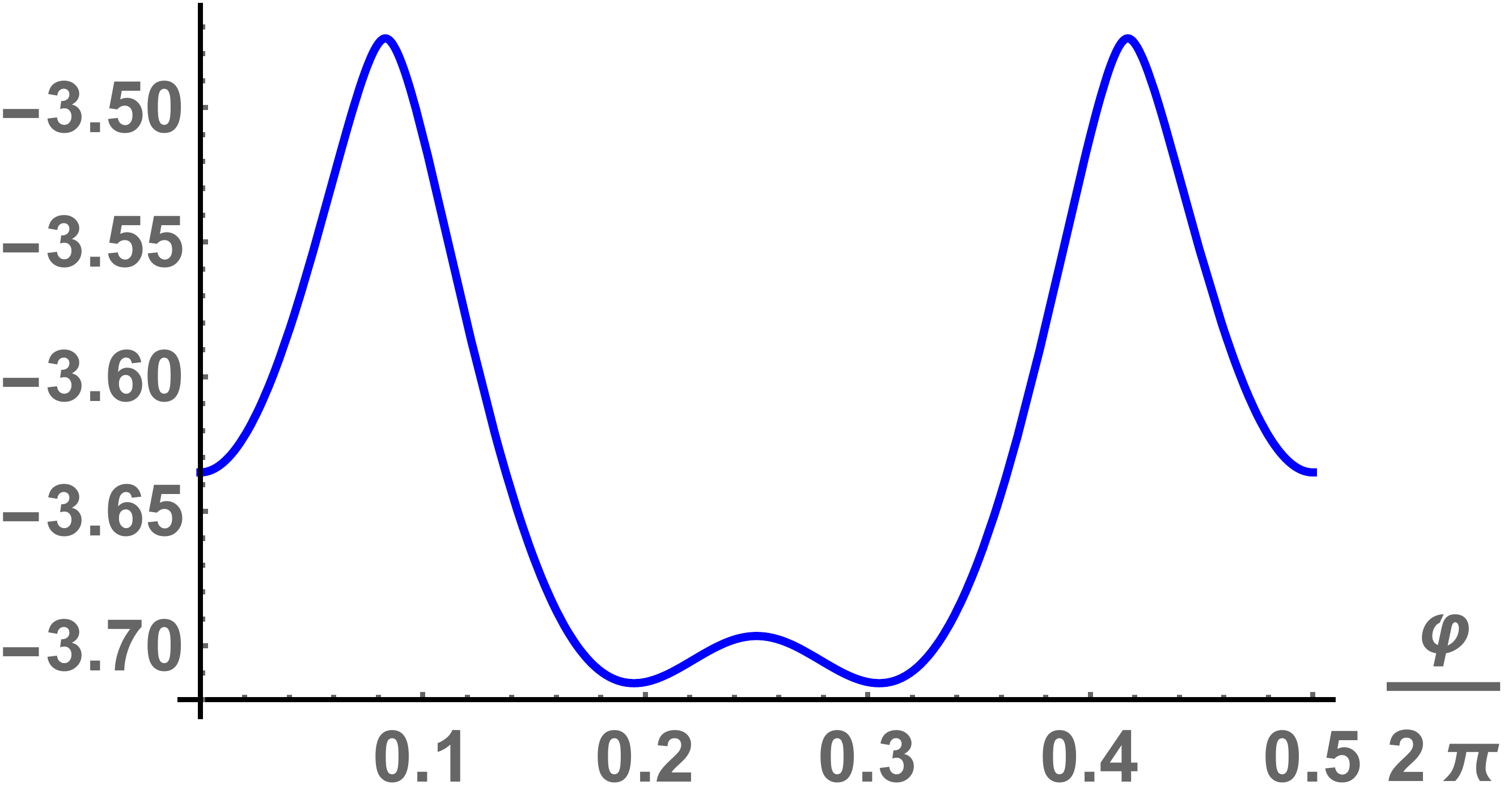}\\ e)
                  }
    \end{minipage}
      \begin{minipage}[h]{.5\linewidth}
\center{
          \includegraphics[width=\linewidth]{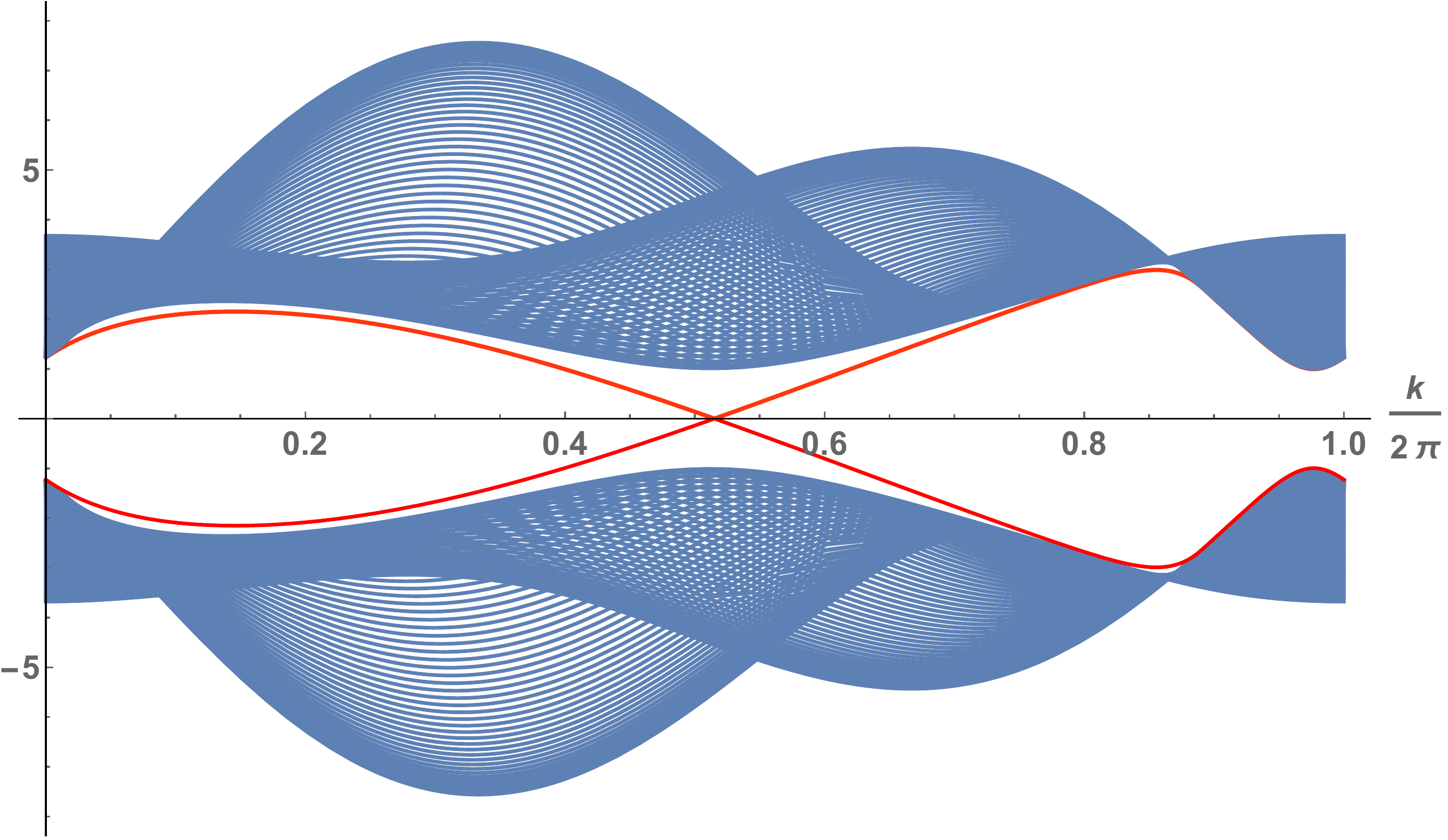}\\ f)
                  }
    \end{minipage}
    \caption{(Color online) Symmetric case $h=0$.
   The energy per cell as a function of the unknown phase difference $\varphi$ at $\tau=0.2$ a), $\tau=1.067$ c) and $\tau=1.3$ e).  The fermion spectrum with boundaries in the TI state calculated for system with zig-zag boundaries at $\tau =0.2$ b),  $\tau=1.067$ d) and $\tau =1.3$ f), the wave vector along the boundary. In the TI state the chiral gapless edge modes  connect the lower and upper bands and intersect at $k_x = \pi$ b) and d),
  the point of the crossing is shifted from $k =\pi$  at $\tau>\tau_0$ f).
       }
    \label{fig:1}
\end{figure}

The proposed model defined on a hexagonal lattice is shown in Fig.\ref{fig:0}. The spectrum of the fermions includes two branches of excitations $\epsilon_{1,2}(\mathbf{k})$, that define the ground-state energy
$E(\{\varphi \})=\sum_{j=1,2}\sum_{\mathbf{k},\epsilon_j(\mathbf{k})<\epsilon_F} \epsilon_j(\mathbf{k})$ ($\epsilon_F$ is the Fermi energy). The phase state of the system is defined by stable solutions for the unknown phases.

The energy of the system is an $2\pi$-periodic even function of $\{\varphi \}$. From symmetry of the sublattices $A\to B$ and $h\to-h$ the solutions satisfy the following relations $\{\varphi_A \}=\{ \pm(\pi+\varphi_B) \}$.
From numerical calculations it follows that the stable solutions for the phase differences corresponding to the minima of the energy meet an insulator state at a half-filling with the following relations for the unknown phases $\varphi_{A}^X=\varphi_{A}^Y=-\varphi_{A}^Z$ and $\varphi_{B}^X=\varphi_{B}^Y=-\varphi_{B}^Z$.
We will show, that at half-filling the ground-state of the model is TI or band insulator for arbitrary $h$ and $\tau \neq 0$, a metal phase is not realized. A gap opening in the bulk spectrum decreases the energy of system in an insulator state.

\emph{Symmetric case h=0}

In the simples case $h=0$ at $\tau<\tau_0=1.067$ the solutions for the phase differences, that correspond to the minimums of the energy of the system, define the symmetric case in the Haldane model \cite{Hal, K1, K2}: $\varphi_A^X =  \varphi_A^Y =-\varphi_A^Z =\pm \frac{\pi}{2}$, $\varphi_B^X =\varphi_B^Y =-\varphi_B^Z =\mp \frac{\pi}{2}$.
At $t>t_0$ the solutions for $\varphi_{A,B}^\beta$ satisfy the following relation $\varphi_{A}^Z=\pi +\varphi_{B}^Z$, therefore the energy of the system depends on one unknown variable $\varphi =\varphi_B^Z$. The energy is a periodic function of $\varphi$ having periodicity $2\pi$. We have illustrated the behavior of the energy of the system as a function of the phase $\varphi$ at different values of $\tau$ (0.2,$\tau_0$, 1.3) in Figs \ref{fig:1} a),c),e).
The Chern number is equal to one, the spectrum has a traditional form in the Haldane model \cite{Hal,K1,K2} (see in Fig.\ref{fig:1} b),d),f)). A nonzero value of the Chern number is a signature of the nontrivial topological
properties of the system, it is an integer topological number which is well defined in insulator state for the band isolated from all other bands. In the TI state the chiral edge modes, that populate
the bulk gap, are located on opposite edges, cross the gap connecting the lower and upper bands and intersect at $k = \pi$ at $\tau \leq \tau_0$ (see in Figs \ref{fig:1} b),d)). At $\tau>\tau_0$ the point of the crossing is shifted from $k =\pi$ f) (the wave vector along the zig-zag boundary), the inversion symmetry is breaken spontaneously.
Each of the states having the energy inside the bulk gap is spatially localized near one of the two edges of the system.

In an exotic case, when $\tau>1$ (the phase diagram of the Haldane model \cite{Hal} is obtained in the case $\tau <1/3$ \cite{K2}), time reversal and inversion symmetries are broken. In the  point $\tau_0 = 1.067$ the structure of the stable solutions of unknown phases is changed: the symmetric case of the Haldane model is realized at $\tau<\tau_0$, at $\tau>\tau_0$  two new solutions for the phases correspond to double degenerated ground state of TI, while the maximum of the energy is reached at the symmetric case. A $C_{3v}$ symmetry of the hexagonal lattice is broken, a $C_{3}$ symmetry with two nonequivalent sublattices is realized when the second-nearest neighbor coupling coefficient is larger than the nearest neighbor coupling.

\emph{Asymmetric case $h \neq 0$}
\begin{figure}[tp]
    \centering{\leavevmode}
\begin{minipage}[h]{.4\linewidth}
\center{
            \includegraphics[width=\linewidth]{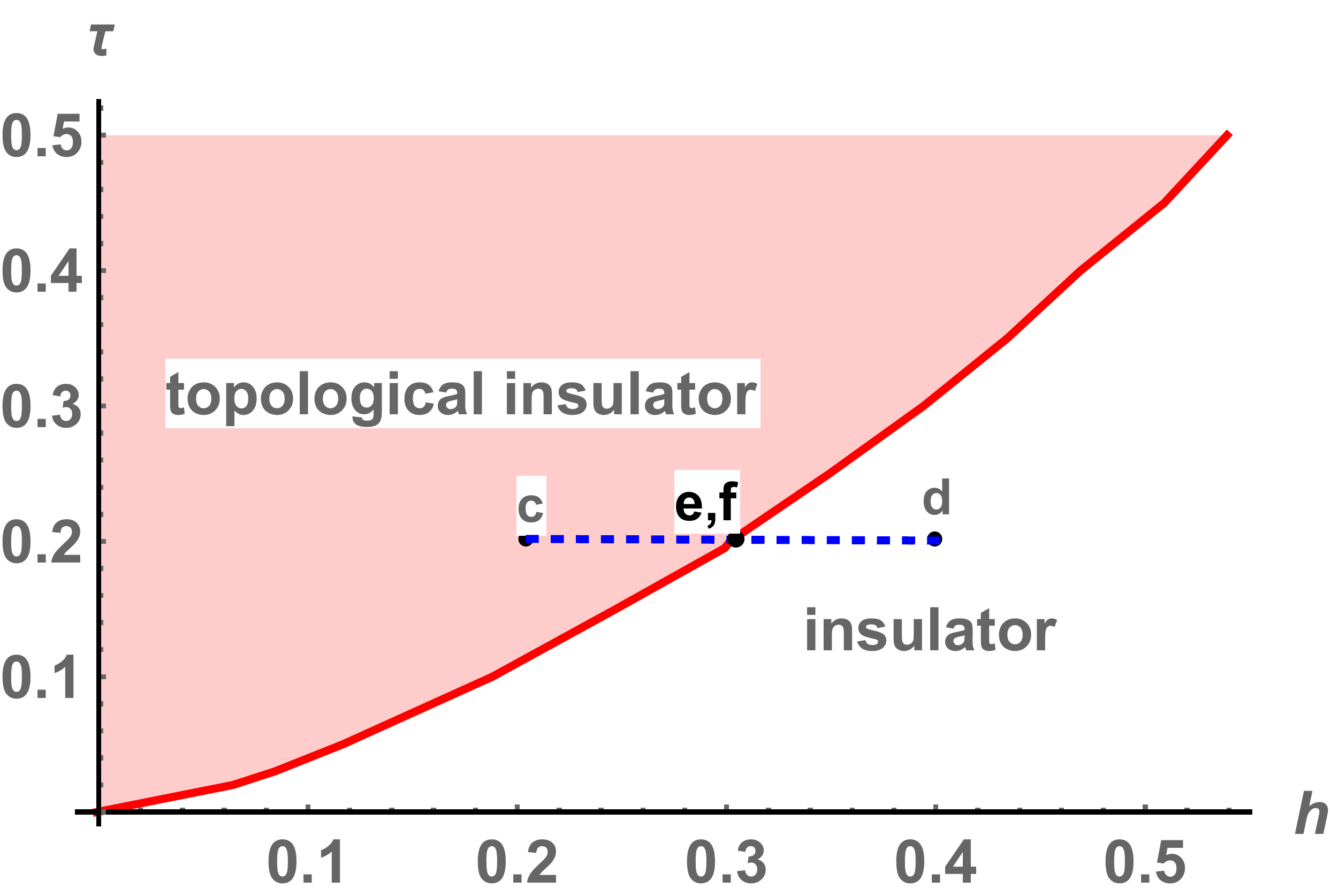} \\ a)
                  }
    \end{minipage}
     \centering{\leavevmode}
\begin{minipage}[h]{.4\linewidth}
\center{
         \includegraphics[width=\linewidth]{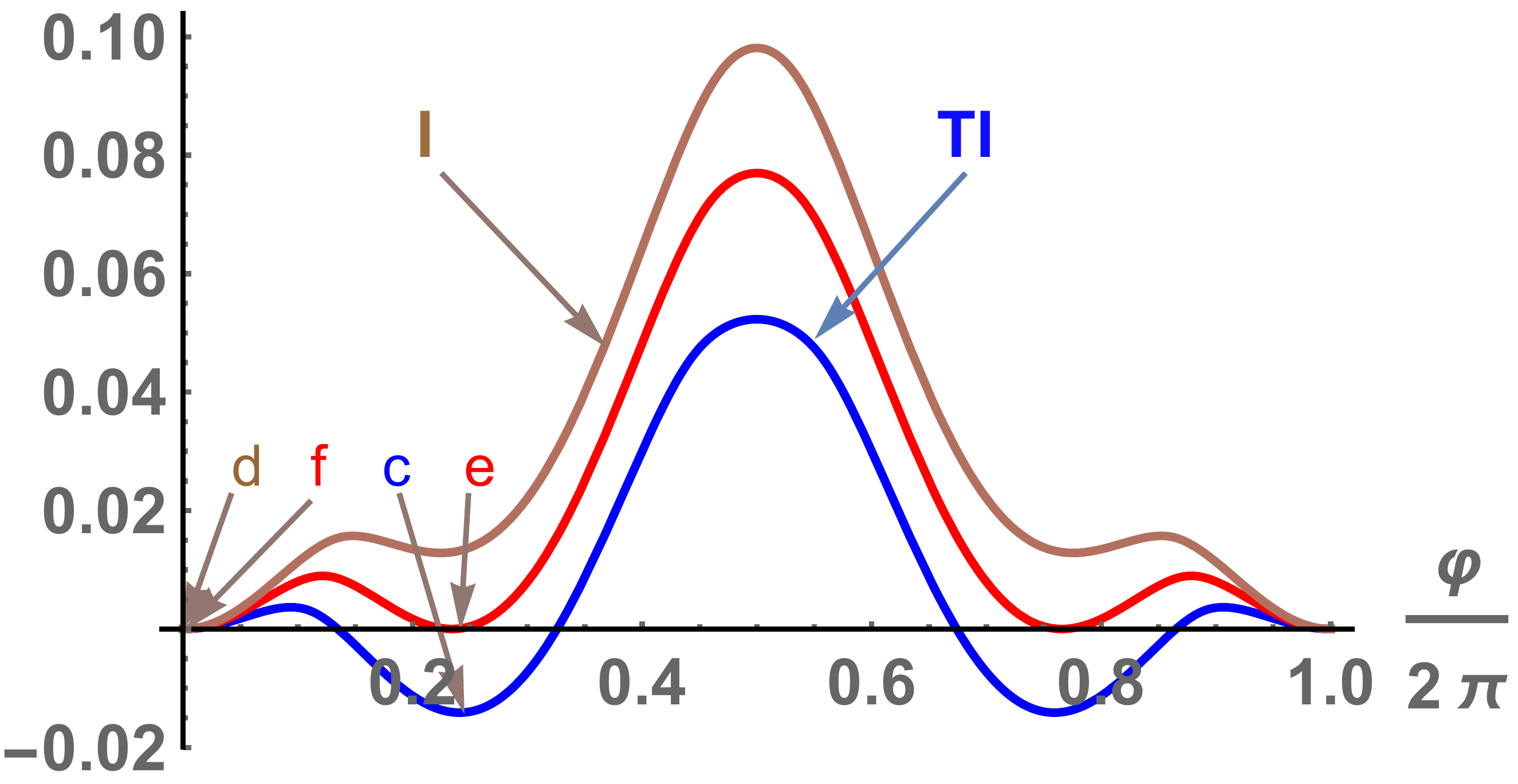} \\ b)
                  }
    \end{minipage}
    \begin{minipage}[h]{.4\linewidth}
\center{
           \includegraphics[width=\linewidth]{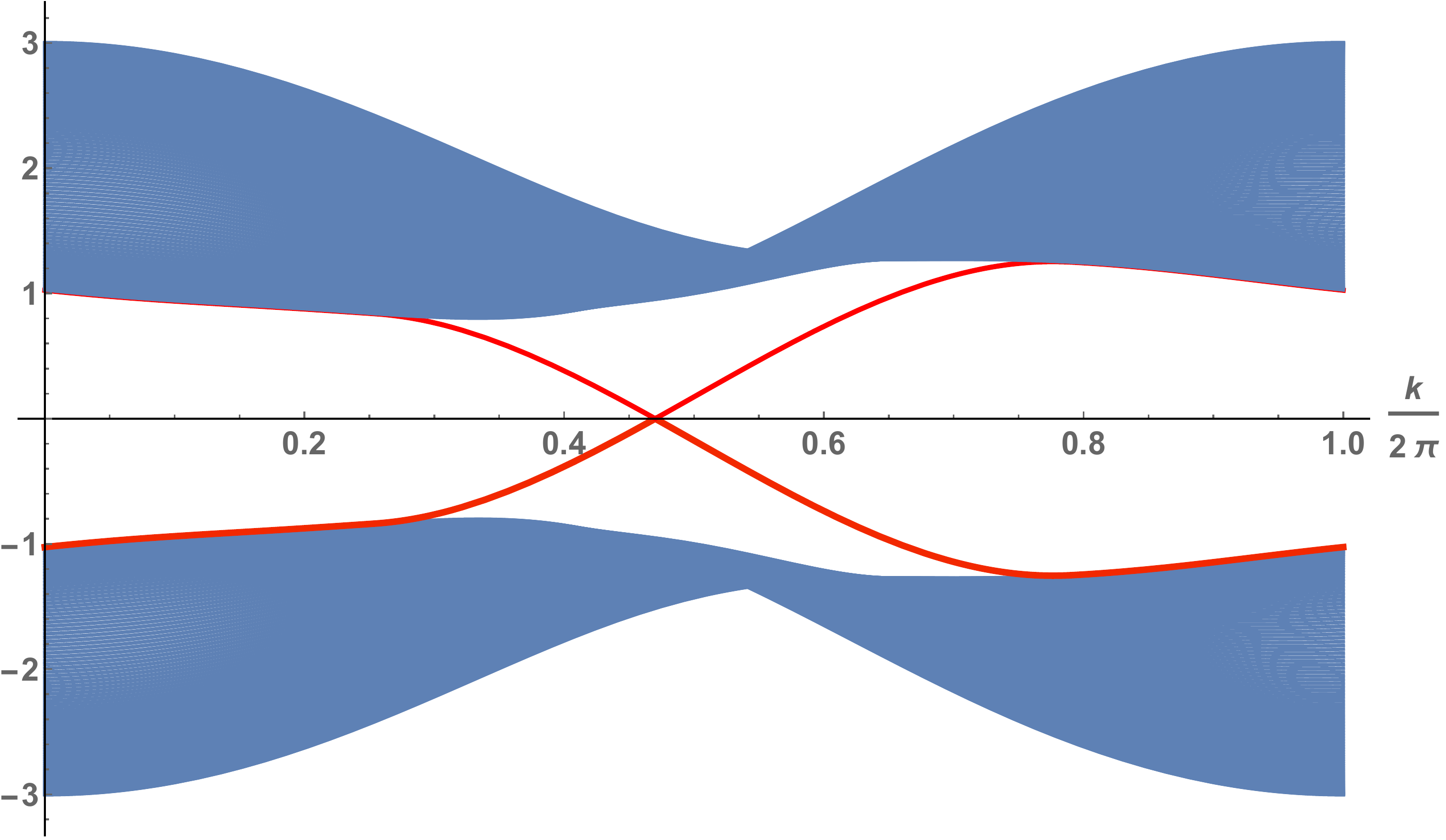} \\ c)
                  }
    \end{minipage}
    \begin{minipage}[h]{.4\linewidth}
\center{
           \includegraphics[width=\linewidth]{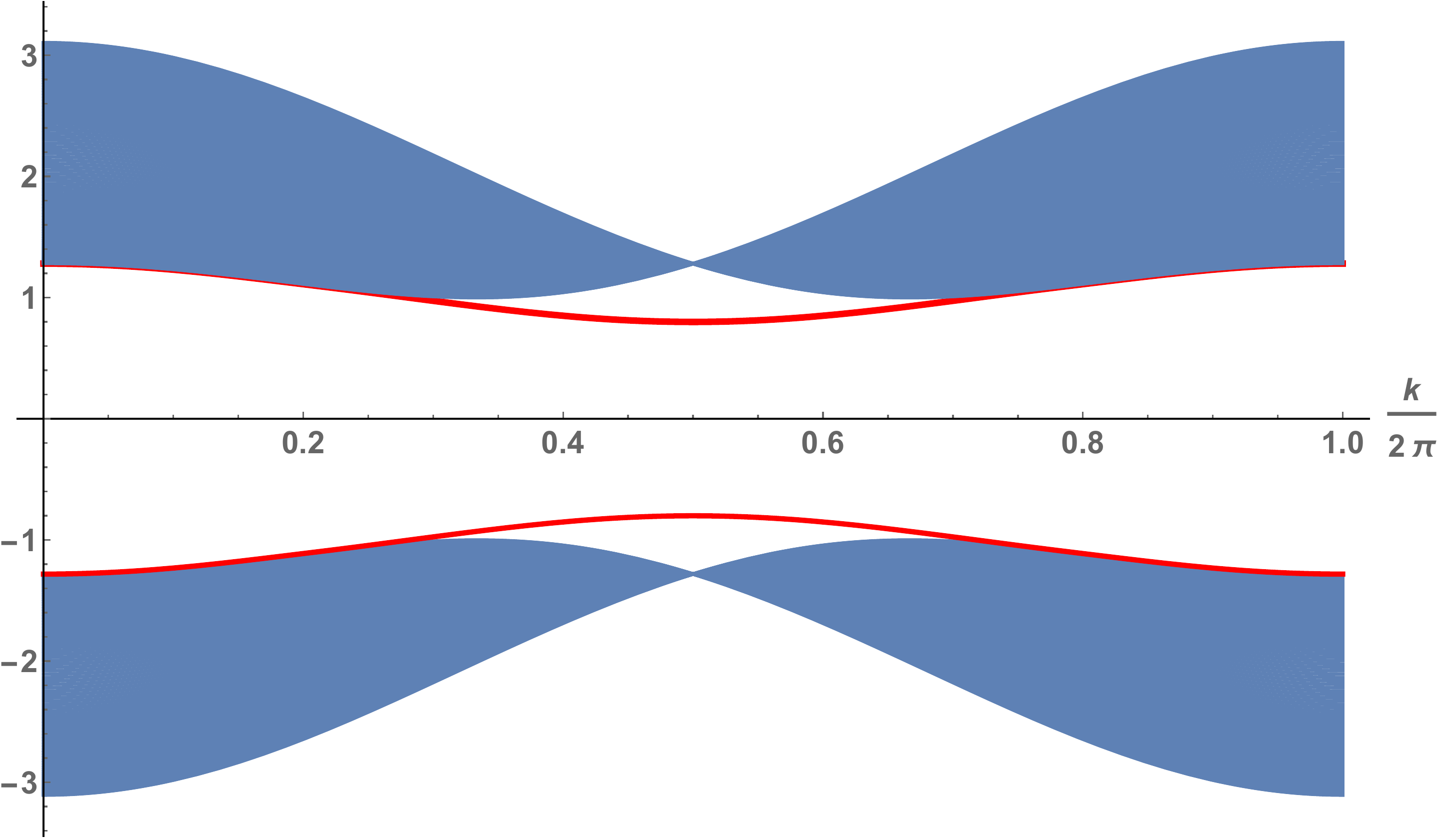} \\ d)
                  }
    \end{minipage}
     \begin{minipage}[h]{.4\linewidth}
\center{
          \includegraphics[width=\linewidth]{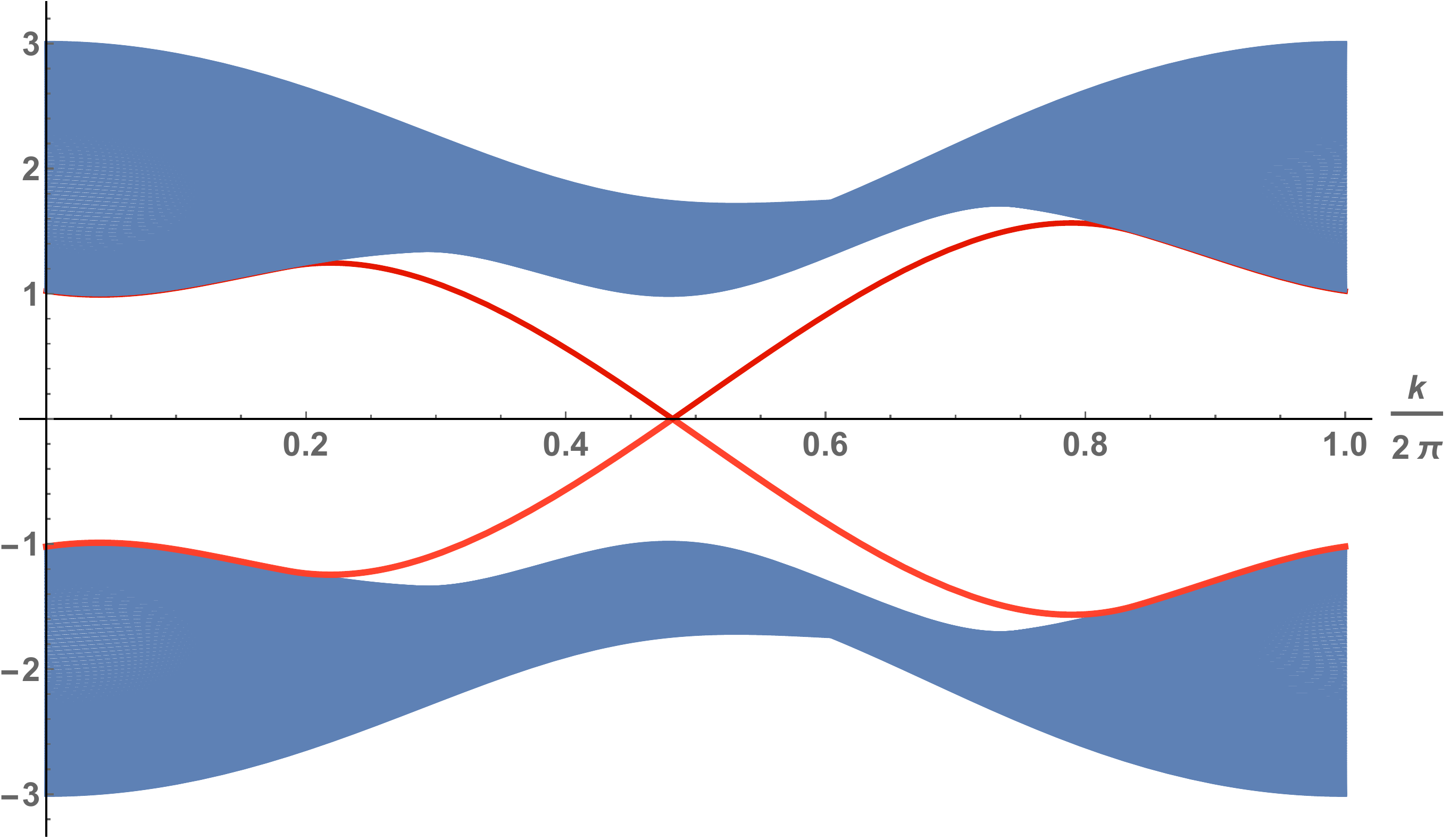} \\ e)
                  }
    \end{minipage}
     \begin{minipage}[h]{.4\linewidth}
\center{
           \includegraphics[width=\linewidth]{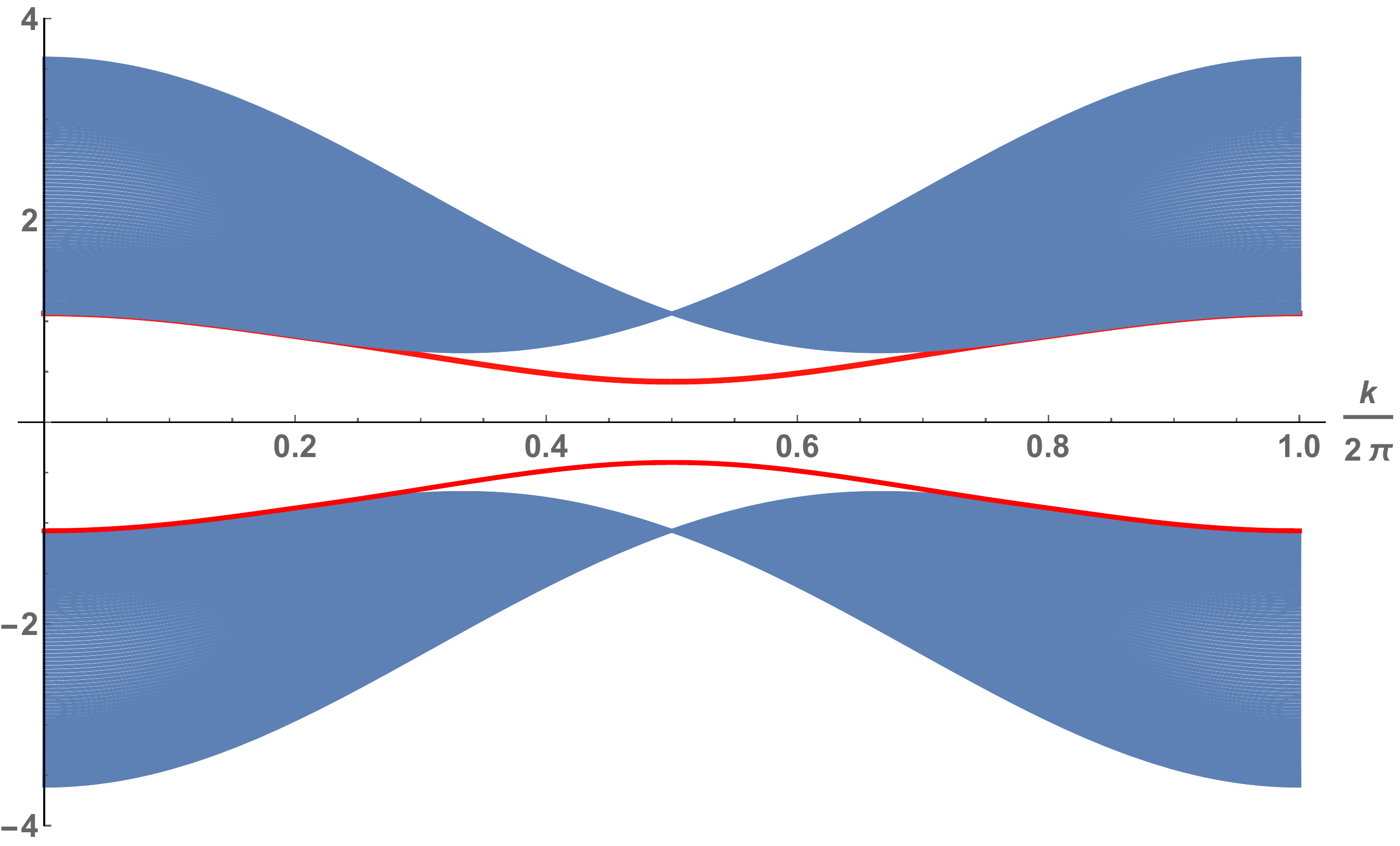} \\ f)
                  }
    \end{minipage}
    \caption{(Color online) Asymmetric case $h\neq 0$. The ground-state phase diagram in the coordinates ($h, \tau$) a), the energy difference $E(\varphi)-E(0)$ defines stable region of existence of TI and band insulator (I) b).  The band structure calculated for zig-zag boundaries in the TI state at $h=0.2, \tau=0.2$ c), band insulator at $h=0.4, \tau=0.2$ d) and at the point of the phase transition at $h=0.3025$ in TI e) and insulator f) states.
       }
    \label{fig:2}
\end{figure}
The staggered potential destroys the symmetry of the spectrum and leads to the phase transition from TI to band insulator.   The following solutions for the phase differences $\varphi_{A}^\beta =\pi, \varphi_{B}^\beta =0$ or $\varphi_{B}^\beta =\pi, \varphi_{A}^\beta =0$ correspond to the band insulator state.  The staggered potential slightly deforms the solutions for $\varphi_{A,B}^\beta$ in the TI state and then the system changes the phase state (see in Fig.\ref{fig:2} b)), because the strongly deformed solutions for the phase differences are unstable.
The ground-state phase diagram is calculated for stable trivial and nontrivial solutions of the phase differences (see in Fig.\ref{fig:2} a)), these solutions are defined by the energy difference $E(\varphi)-E(0)$. Typical behavior of the energy $E(\varphi)-E(0)$  at $h=0.2,\tau =0.2$ in the TI state,  at $h=0.4,\tau =0.2$ in insulator state and $h=0.3025,\tau =0.2$ in the point of the phase transition is shown in Fig. \ref{fig:2} b). The topological phase transition separates the states of TI and  band isolator, the phase states with different static configurations of stable phases and the Chern numbers. The Figs \ref{fig:1} c)-f) illustrate the evolution of the spectrum of the system along line $\tau=0.2, 0.2\leq h \leq 0.4$ from TI state at $ h=0.2$ f)  via the point of the topological phase transition at $h=0.3025$ e),f) to band insulator  at $h=0.4$ d). In the insulator phases the bulk gaps are not closed at the point of the phase transition (see in Figs \ref{fig:1} e),f))(because these phase states with different static configurations of the phases), the structure of the edge modes is changed cardinally, that indicates the changing of the Chern number.

The Chern number, equal to one, defines the nontrivial topology of the system. The spectrum of TI includes the chiral edge modes that populate the gap and merge with the bulk states, the point of the crossing is shifted from $k =\pi$. The gapless edge modes and the Chern number define the charge Hall conductance in TI (see in Figs \ref{fig:2} c).

\section{Conclusion }

In conclusion, in this work we have derived a new approach for description of TIs.
We have shown that due to nontrivial stable solutions for the phases, that define the hopping integrals of the spinless fermions along different directions of the link, the time-reversal symmetry is spontaneously broken.
These solutions are define static configurations of the phases, reveal the nontrivial topological properties of the system in the insulator state. In this context, we have also shown that TI is stable in a wide region  of values of the  coupling parameters, namely the next-nearest neighbor hopping integrals and the staggered potential. We have studied the Chern numbers, the chiral gapless edge modes in TI. Depending on the relation between parameters of the model, band insulator or TI are realized in the system.
In contrast to traditional approach to TI,  we have proposed the mechanism of spontaneous breaking of time-reversal symmetry in the result of which the TI state is realized in an absence of any interaction, that breaks the time-reversal symmetry. We have shown also, that the Haldane model can be implemented in real compounds of the condensed matter physics. This consideration is reasonable for charge neutral particles (such as fermions in cold atom system), or for electrons on superlattices (such as twisted bilayer graphene).

\section*{References}
\bibliography{mybibfile}

\end{document}